\documentclass[twocolumn,tighten]{aastex63}
\usepackage{amssymb, amsmath,framed}

\usepackage{bm}
\expandafter\ifx\csname package@font\endcsname\relax\else
 \expandafter\expandafter
 \expandafter\usepackage
 \expandafter\expandafter
 \expandafter{\csname package@font\endcsname}
\fi
\def\bq{\begin{equation}}
\def\eq{\end{equation}}
\def\bqy{\begin{eqnarray}}
\def\eqy{\end{eqnarray}}
\def\p{\partial}

\def\p{\partial}

\def\R{\mathbb{R}}

 \def\q#1#2{q^{\hspace{1pt}#1}_{\,,\hspace{1pt}#2}}
 \def\a#1#2{a^{\hspace{1pt}#1}_{\,,\hspace{1pt}#2}}
 \def\d#1#2{\delta^{\hspace{1pt}#1}_{\,,\hspace{1pt}#2}}

\begin{document}
\title{\large{Implications of Abiotic Oxygen Buildup for Earth-like Complex Life}}

\correspondingauthor{Manasvi Lingam}
\email{mlingam@fit.edu}

\author{Manasvi Lingam}
\affiliation{Department of Aerospace, Physics and Space Sciences, Florida Institute of Technology, Melbourne FL 32901, USA}
\affiliation{Institute for Theory and Computation, Harvard University, Cambridge MA 02138, USA}

\begin{abstract}
One of the chief paradoxes of molecular oxygen (O$_2$) is that it is an essential requirement for multicellular eukaryotes on Earth while simultaneously posing a threat to their survival via the formation of reactive oxygen species. In this paper, the constraints imposed by O$_2$ on Earth-like complex life are invoked to explore whether worlds with abiotic O$_2$ inventories can harbor such organisms. By taking the major O$_2$ sources and sinks of Earth-like planets into account using a simple model, it is suggested that worlds that receive time-averaged X-ray and extreme ultraviolet fluxes that are $\gtrsim 10$ times higher than Earth might not be capable of hosting complex lifeforms because the photolysis of molecules such as water may lead to significant O$_2$ buildup. Methods for testing this hypothesis by searching for anticorrelations between biosignatures and indicators of abiotic O$_2$ atmospheres are described. In the event, however, that life successfully adapts to high-oxygen environments, these worlds could permit the evolution of large and complex organisms. \\
\end{abstract}

\section{Introduction} \label{SecIntro}
The evolution of oxygenic photosynthesis and the subsequent rise in the inventories of atmospheric and oceanic molecular oxygen (O$_2$) is known to have caused a dramatic transformation of Earth's biosphere \citep{Knoll15}. Among other factors, the rise in O$_2$ levels is believed to have facilitated the emergence of multicellular eukaryotic life on Earth \citep{Nur59}. In a similar vein, it has been argued that O$_2$ ranks high among the list of desiderata for complex multicellularity to arise elsewhere \citep{CGZM,WSH19}. 

However, the presence of O$_2$ is a double-edged sword, seeing as how complex multicellular organisms cannot live without it, but the very same gas poses an incessant threat to their existence. It is for this reason that this issue has been dubbed the ``oxygen paradox'' \citep{Dav95}. The threat posed by O$_2$ is a direct consequence of its chemical properties, namely, the fact that it has two unpaired electrons. Cells offer plenty of pathways by which molecular oxygen can be reduced to yield highly reactive intermediates, which in turn are responsible for oxygen toxicity \citep{HG15}. Thus, on the one hand, O$_2$ seems to be an essential prerequisite for complex life, but on the other, high concentrations of O$_2$ are potentially lethal to these organisms. 

Hence, it is apparent that a careful analysis of the atmospheric O$_2$ inventories of other worlds is warranted from a biological standpoint. This issue acquires an additional significance when recent developments in exoplanetary science are duly taken into account. In particular, M-dwarf exoplanets have attracted a great deal of attention from both theoretical and observational standpoints, partly due to their commonality and partly because of the discovery of several high-profile exoplanetary systems in recent times \citep{LL19}. One of the most notable among them has to do with the high X-ray and extreme ultraviolet (XUV) fluxes received by these planets \citep{Lin19}. 

A number of models suggest that high XUV fluxes play a vital role in the buildup of massive O$_2$ atmospheres by way of desiccation and photolysis \citep{Kas88,WP14,RK14,LB15,BSO17,PSD}. However, other models indicate that this buildup may be mitigated by a number of mechanisms that may result in modest O$_2$ accumulation and H$_2$O retention \citep{Ti15,WSF18,HFH18,John19}. At this juncture, it is essential to recognize that the buildup of dense abiotic O$_2$ atmospheres is often (but not always) accompanied by the depletion of surface water, because it serves as the source for oxygen by undergoing evaporation and photolysis by XUV radiation. Hence, the resultant desiccation could also impose stringent constraints on their habitability, although this issue depends on the initial water inventories of these worlds \citep{TGJ18}.

Hence, in the event that the buildup of massive O$_2$ atmospheres is rendered feasible by abiotic means, there are two phenomena that merit consideration. First, the requirements for Earth-like complex life can be assessed by drawing upon the available data for oxygen toxicity. Second, if life does find a way to adapt to the myriad challenges posed by hyperoxic environments, the high O$_2$ concentrations become a boon of sorts and permit the evolution of large organisms, although high levels of O$_2$ could fuel fires that may engender other difficulties for life. To some degree, the trend toward larger organismal size has been observed in the Permo-Carboniferous period, where elevated O$_2$ levels - roughly $70\%$ higher than the present-day value \citep{GA95} - exhibit a strong positive correlation with an increase in maximal insect sizes \citep{HKV10,CK12}. Insect gigantism was prevalent in this epoch with dragonflies of genus \emph{Meganeura} reaching wingspans of $0.7$ m \citep{TL05,SPH16}. 

It is therefore the goal of this paper to explore the above two issues. In Sec. \ref{SecOTox}, we provide a primer on oxygen toxicity and discuss the threshold(s) at which O$_2$ becomes lethal to Earth-based complex multicellular organisms. Next, we develop a simple model in Sec. \ref{SecOSS} that takes the major O$_2$ sources and sinks into account to assess the partial pressure of atmospheric oxygen. This is followed by a presentation of our central results in Sec. \ref{SecPros}. We conclude with a summary of our findings and their accompanying implications in Sec. \ref{SecConc}.

\section{Oxygen Toxicity: A Synopsis}\label{SecOTox}
By the 1960s, it was already well established that high concentrations of O$_2$ were responsible for the inhibition of metabolic processes in eukaryotes as well as prokaryotes due to the intricate interactions between O$_2$ and a number of proteins such as flavoproteins and ferredoxins \citep{Hau68}. Subsequently, a number of experiments have been conducted to uncover the mechanisms by which O$_2$ can cause extensive damage. 

Of the various model organisms, \emph{Escherichia coli} has been the most widely studied \citep{Im13}. When subjected to high extracellular oxygen concentrations, the diffusion of O$_2$ across cellular membranes is followed by the abstraction of electrons from reduced flavoprotein cofactors, thereby driving the formation of reactive oxygen species (ROS) such as superoxide (O$_2^-$) and hydrogen peroxide (H$_2$O$_2$). The resultant ROS suppress enzymatic activity (especially in the case of metalloenzymes) and damage lipid membranes as well as DNA \citep{Jam89}, thus promoting mutagenesis and impairing organismal growth \citep{Im03}.

Both prokaryotes and eukaryotes are adversely impacted at high concentrations of O$_2$, with the latter typically displaying higher sensitivity to hyperoxia than the former \citep{BS14}. A number of studies have established that microbes such as \emph{Enterococcus faecalis}, \emph{Bacillus subtilis}, \emph{Lactobacillus sake}, \emph{Vibrio vulnificus} and \emph{E. coli} experience inhibition of respiration and growth to varying degrees when the partial pressure of oxygen ($\mathrm{P}_\mathrm{O_2}$) exceeds $\sim 1$ atm \citep{GGF74,BVB76,HSRG,AB01,TBKF,TIS12}. Experiments suggest that $\sim 20$ atm of O$_2$ constitutes a lethal limit for the likes of \emph{Saccharomyces cerevisiae} and \emph{E. coli} \citep{GGF74}. 

The oxygen tolerance levels for animals tend to be lower in general. For example, the exposure of rats to hyperoxic environments comprising $100\%$ O$_2$ for several days caused respiratory failure and death \citep{CBF80,ORS95} and similar effects have been documented for humans \citep{KM13}. Likewise, the exposure of human, Chinese hamster, and mouse cells to oxygen partial pressures of $\sim 1$ atm was accompanied by a strong inhibition of cell growth and extensive DNA damage \citep{LM92,CTL93,Mar95}. Aside from animals, plants are also very susceptible to oxygen toxicity as they undergo cell wall damage, diminished seed viability, and inhibition of chloroplast growth, to name a few \citep{Sch02,GT10}.

At the same time, however, it is important to recognize that a number of defense mechanisms exist to combat the deleterious effects of ROS. Antioxidant compounds such as glutathione, ubiquinone, and the vitamins C and E are selectively oxidized to protect more sensitive components of cells from the same fate \citep{Cad89}. A number of antioxidant enzymes, such as the superoxide dismutases and quinone reductases, catalyze the transformation of oxidants into less reactive species \citep{Frid95,Dav95}. Furthermore, DNA-repair enzymes, phospholipases, and proteasomes collectively act to recognize and eliminate cellular components that have been oxidized \citep{PHF06}.

As the preceding discussion illustrates, the identification of ``universal'' limits for oxygen toxicity is rendered complex due to the fact that physiological damage is dependent not only on the partial pressure of O$_2$ but also the exposure time and the rich palette of antioxidant mechanisms at play. With these caveats in mind, the threshold for complex multicellular life (after sufficiently long exposure) on Earth appears to be $\mathrm{P}_c \sim 1$-$10$ atm \citep[Section 1.5]{HG15}. Hence, this range is adopted henceforth in our analysis, albeit with the explicit acknowledgement that our analysis is indubitably parochial by virtue of its Earth-centric focus. 

Although parochial in nature, studies of this nature are nonetheless advantageous because they permit us to make concrete and testable predictions and could play a valuable role in the selection of target planets. It is for this reason that Earth-based limits are routinely employed in astrobiological studies, with the recent formulation of the habitable zone (HZ) for complex life based on CO$_2$ and CO constraints \citep{SRO19} constituting a compelling example of this approach.

\section{Oxygen sources and sinks on Earth-like planets}\label{SecOSS}
In keeping with the prior outlook, we will restrict ourselves to worlds that are near-exact geological analogs of modern Earth. There are a number of sources and sinks for O$_2$ that must be taken into consideration. As we wish to study the consequences of \emph{abiotic} O$_2$ buildup, we will set aside the contribution arising from organic carbon burial (i.e., involving photosynthesis). 

We begin by quantifying the various sources of O$_2$. Aside from organic carbon, the burial of pyrite and Fe$^{2+}$ acts as an effective source. The net production of O$_2$ via this pathway is estimated using \citet[Table 10.1]{CK17}, which yields
\begin{equation}\label{O2bur}
    \dot{N}_\mathrm{O_2} \approx 3.6 \times 10^{-2}\,\mathrm{bar/Myr}\,\left(\frac{R}{R_\oplus}\right)^2
\end{equation}
where $R$ denotes the radius of the world. In the above expression as well as subsequent equations, the $R^2$ scaling enters because the flux is held constant for analogs of modern Earth \citep{MaLo19}. Here, we have expressed our result in units of bar/Myr as it will be convenient shortly hereafter. The other major source term for O$_2$ entails the evaporation of water followed by H$_2$O photolysis and hydrogen escape, thereby facilitating the buildup of atmospheric O$_2$ \citep{LB15}. The net production of O$_2$ through this channel is determined by synthesizing two expressions \citep[Equations 11-13]{LL19} as follows:
\begin{eqnarray}\label{O2Phot}
&& \dot{N}_\mathrm{O_2} \approx 0.138\,\mathrm{bar/Myr}\,\left(\frac{F_\mathrm{XUV}}{F_\oplus}\right)\left(\frac{R}{R_\oplus}\right)^{-1}\Theta(F_c - F_\mathrm{XUV}) \nonumber \\
&& \hspace{0.5in} + \, 5.35\,\mathrm{bar/Myr}\,\left(\frac{R}{R_\oplus}\right)^{3.4}\Theta(F_\mathrm{XUV}-F_c),
\end{eqnarray}
where $\mathcal{F}_\oplus \approx 4.6 \times 10^{-3}\,\mathrm{J\,m^{-2}\,s^{-1}}$ denotes the average XUV flux incident on Earth and $\Theta$ represents the Heaviside function. Here, we have adopted an XUV absorption efficiency of $0.3$ \citep{LB15,BSO17} and utilized the mass-radius relationship $M \propto R^{3.7}$ \citep{ZSJ}. The critical XUV flux ($F_c$) where the change in scaling occurs is defined as
\begin{equation}
    F_c = 38.8\,F_\oplus\,\left(\frac{R}{R_\oplus}\right)^{4.4}.
\end{equation}

Next, we turn our attention to the various O$_2$ sinks. The first major sink worth noting entails contintental weathering, which is further decomposed into contributions from carbon, sulfide and Fe$^{2+}$ weathering. From \citet[Table 10.2]{CK17}, the ensuing contribution to the total O$_2$ budget is
\begin{equation}
    \dot{N}_\mathrm{O_2} \approx -7.5 \times 10^{-2}\,\mathrm{bar/Myr}\,\left(\frac{R}{R_\oplus}\right)^2,
\end{equation}
where the negative sign on the RHS signifies the presence of an O$_2$ sink. Aside from continental weathering, the depletion of O$_2$ occurs due to reactions with reducing gases from volcanism, serpentinization and metamorphism. As per \citet[Table 10.3]{CK17}, the contribution from this sink is expressible as
\begin{equation}
    \dot{N}_\mathrm{O_2} \approx -3.5 \times 10^{-2}\,\mathrm{bar/Myr}\,\left(\frac{R}{R_\oplus}\right)^2.
\end{equation}
Another major sink, often overlooked in most studies, entails the non-thermal escape of O$^+$ ions due to stellar wind stripping \citep{Lam13,LL19}. Although the complexity of this phenomenon requires numerical modeling using multi-fluid magnetohydrodynamics \citep{GGD17,DLMC,DHL17,DJL18,Aira19}, we will adopt the simple analytic prescription delineated in \citet{AGK17} instead. If the O$^+$ ions are assumed to be sourced from oxygen \citep[Table 1]{DHL19}, by using \citet[Equation 1]{AGK17}, we have
\begin{equation}\label{O2esc}
    \dot{N}_\mathrm{O_2} \approx -4.6 \times 10^{-4}\,\mathrm{bar/Myr}\,\left(\frac{F_\mathrm{XUV}}{F_\oplus}\right)\left(\frac{R}{R_\oplus}\right)^2.
\end{equation}
The above sink exhibits the same functional dependence on $R$ and $F_\mathrm{XUV}$ as that of the photochemical escape of hot O \citep{CRF17}, but the latter appears to constitute a smaller sink in comparison across the world's entire history \citep{DLM18}, owing to which it may be neglected herein.

In principle, it is straightforward to estimate the total amount of atmospheric O$_2$ that is generated over a given time interval ($\Delta t$) by applying the following formula. 
\begin{equation}\label{TotO2gen}
 \mathrm{P}_\mathrm{O_2} = \int_0^{\Delta t} \dot{N}_\mathrm{O_2}^{(\mathrm{tot})}\,\mathrm{dt},
\end{equation}
where $\dot{N}_\mathrm{O_2}^{(\mathrm{tot})} = \sum_i \dot{N}_\mathrm{O_2}$ is obtained by summing (\ref{O2bur})-(\ref{O2esc}) accordingly. At this stage, we encounter an immediate difficulty because all of the source and sink terms are time-dependent functions. In consequence, a rigorous calculation of the above integral is not feasible because of the many unknowns involved ranging from XUV evolution \citep{Ti15} to the water inventory and redox history of the Earth-analog. Hence, we introduce the simplifying assumption that the variables in the source and sink terms are roughly held constant at their present-day measured values.\footnote{Based on the fact that $\dot{N}_\mathrm{O_2}^{(\mathrm{tot})}$ is a linear function of $F_\mathrm{XUV}$, we can interpret the latter quantity henceforth as the temporal average of the XUV flux over the age of the world.} As explained in Sec. \ref{SecOTox}, the criterion $\mathrm{P}_\mathrm{O_2} < \mathrm{P}_c$ is necessary for the sustenance of Earth-like complex life, which implies that 
\begin{equation}\label{InEqO2}
\dot{N}_\mathrm{O_2}^{(\mathrm{tot})} \Delta t < \mathrm{P}_c,
\end{equation}
where the integral in (\ref{TotO2gen}) has been eliminated on account of the prior assumption. By using this inequality, we can gauge the parameter space wherein the existence of Earth-like complex life might be feasible.

Before doing so, a couple of important observations are in order. First, we wish to reiterate that the most dominant source of O$_2$, corresponding to (\ref{O2Phot}), is concomitantly accompanied by the loss of H$_2$O. In fact, the depletion of $\sim 270$ bars of H$_2$O - which is equivalent to the production of $\sim 135$ bars of O$_2$ assuming that the stoichiometric ratio is preserved - would result in the complete loss of Earth's oceans. By inspecting (\ref{O2Phot}), we see that the depletion of many times the water inventory of Earth's oceans is conceivable at high enough XUV fluxes. It must be noted, however, that this does not translate to a death knell for the habitability of all worlds because some of them may have sufficiently high water inventories to avoid total desiccation \citep{Kuc03,LSS04,KF18}.

Second, we have implicitly assumed that $\mathrm{P}_c$ is static through time in (\ref{InEqO2}), although we did highlight the temporal dependence of $\dot{N}_\mathrm{O_2}^{(\mathrm{tot})}$. In actuality, a time-independent choice of $\mathrm{P}_c$ constitutes an oversimplification, because the threshold for multicellular eukaryotes is not necessarily the same as that of microbes (especially anaerobic ones) and prebiotic compounds. For example, in the case of the latter, the synthesis of prebiotic compounds, and potentially the origin of life, is believed to have required a reducing atmosphere \citep{McCo13,PLL16}; to put it differently, the synthesis of biomolecules may be suppressed in massive O$_2$ atmospheres as they could undergo rapid oxidation. Moreover, putative organisms might have to compete with O$_2$ to access the free energy from redox reactions with organic molecules \citep[Section 6.3]{LB15}. The amount of energy to synthesize the same quantity of biomass is approximately $13$ times higher in oxygenated environments compared to their anoxic counterparts \citep[Table 4]{McA05}.

Therefore, the ensuing analysis is predicated on the premise that microbial life is already existent on the world,\footnote{A potential avenue by which this might occur is through interstellar panspermia \citep{BMM12,LiLo18,GLL18}, although this pathway faces many of its own challenges \citep{Mel03,Nic09}.} because we will restrict ourselves to exploring the prospects for Earth-like complex biota. A more encompassing and realistic study will, however, need to explicitly account for the temporal aspects of evolution and thereby assess whether: (i) life is capable of emerging on worlds with highly oxygenated atmospheres, and (ii) microbial lifeforms could subsequently become more complex over time via eco-evolutionary processes.

\begin{figure*}
$$
\begin{array}{cc}
  \includegraphics[width=7.5cm]{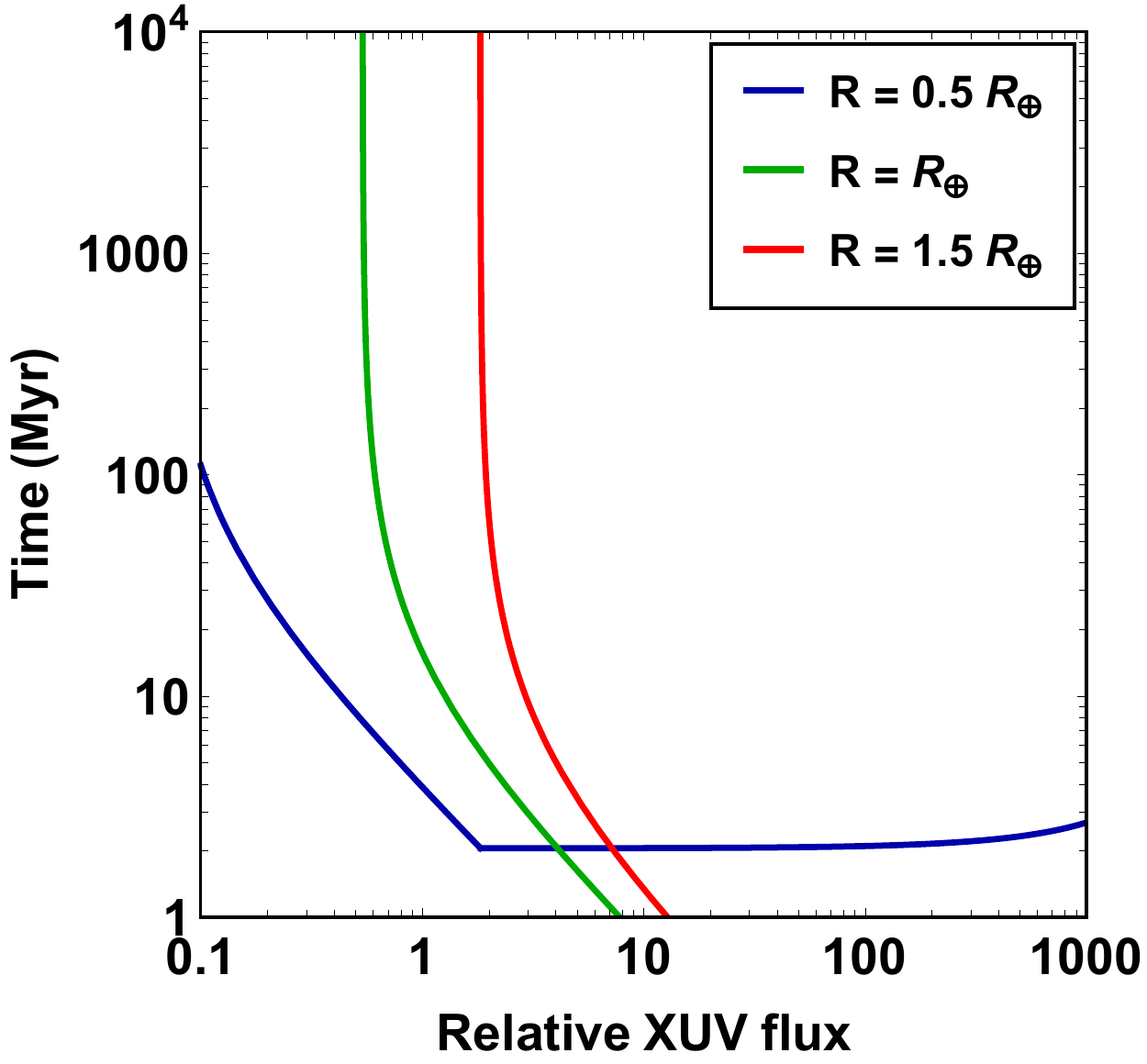} &  \includegraphics[width=7.5cm]{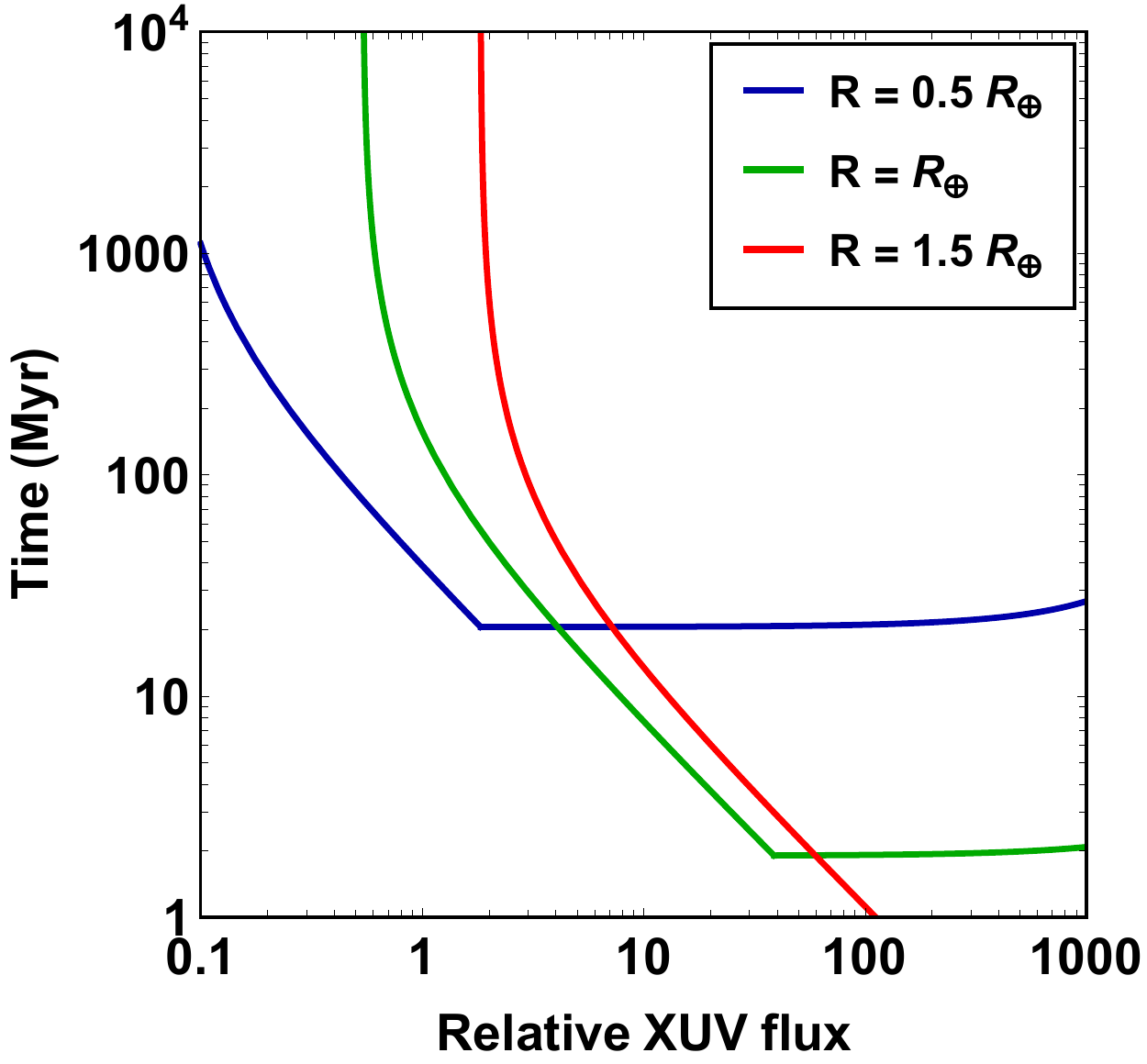}\\
\end{array}
$$
\caption{In both panels, the permitted range of values - namely, the regions below the corresponding curves - for the average XUV flux (relative to modern Earth) and age of the world (in Myr) are illustrated. For the left panel, an oxygen toxicity threshold of $1$ bar was considered whereas the right panel supposes a threshold of $10$ bar. Note that the blue, green and red curves signify worlds with $R = 0.5 R_\oplus$, $R = R_\oplus$ and $R = 1.5 R_\oplus$, respectively.}
\label{FigOxyTox}
\end{figure*}

\begin{figure}
\includegraphics[width=7.5cm]{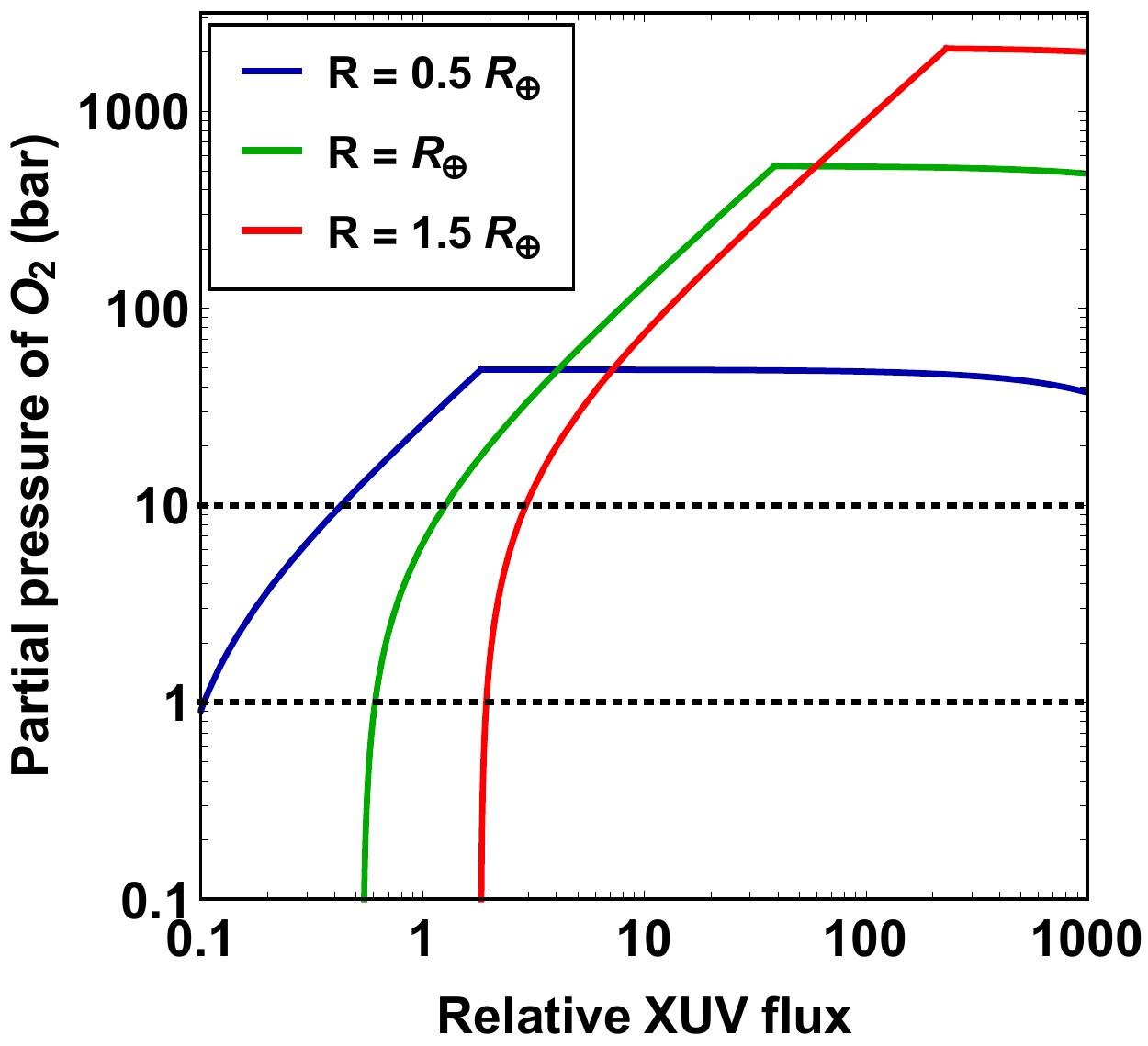} \\
\caption{Partial pressure of O$_2$ (in bars) as a function of the average XUV flux relative to Earth for a world that is $\sim 100$ Myr old. The blue, green and red curves signify worlds with $R = 0.5 R_\oplus$, $R = R_\oplus$ and $R = 1.5 R_\oplus$, respectively. The two dotted horizontal lines represent the putative limits for oxygen toxicity as discussed in Sec. \ref{SecOTox}.}
\label{FigPressO}
\end{figure}

\section{Prospects for Complex Life}\label{SecPros}

The inequality (\ref{InEqO2}) comprises three different variables, namely, $R$, $F_\mathrm{XUV}$ and $\Delta t$ and must be evaluated for two different choices of $\mathrm{P}_c$ (viz. $1$ atm and $10$ atm) for reasons outlined previously in Sec. \ref{SecOTox}.

Our final results are depicted in Fig. \ref{FigOxyTox}. A number of general conclusions can be drawn from this figure. To begin with, for $F_\mathrm{XUV} \lesssim F_\oplus$ and $R \gtrsim R_\oplus$, we find that $\Delta t > 1$ Gyr becomes feasible. In other words, for this class of worlds, the buildup of atmospheric O$_2$ is sufficiently low to avoid the dangers of oxygen toxicity. However, once we enter the regime of $F_\mathrm{XUV} \gtrsim F_\oplus$, we observe that $\Delta t \lesssim 100$ Myr is necessary to avoid the harmful effects of oxygen toxicity; this bound is applicable to all sizes considered herein. Hence, at sufficiently high XUV fluxes, the sustenance of Earth-like complex life is rendered very difficult on most planets. 

A synthesis of paleontological, phylogenetic and theoretical evidence indicates that the origin of life on Earth required at most $\mathcal{O}(100)$ Myr after its formation \citep{DPG17,BPC18,BBB}. On the other hand, the emergence of complex multicellular life, required a few Gyr, although this delay was partly a consequence of low atmospheric O$_2$ \citep{Knoll15}.\footnote{This issue of oxygen paucity is, however, unlikely to be an issue on worlds with substantial abiotic O$_2$.} While these timescales are valid for Earth, we cannot infer much from this single datum \citep{ST12}. Nevertheless, in the absence of further information, we will choose $\Delta t \sim 100$ Myr as a fiducial lower bound on the time required for the origin of complex multicellularity \citep{McK96,LL18}. 

For this choice, we invoke the inequality (\ref{InEqO2}) and plot the ensuing results in Fig. \ref{FigPressO}. The first trend we discern is that larger worlds exhibit lower O$_2$ buildup at smaller values of the XUV flux, but eventually increase rapidly and undergo near-saturation; the saturation thresholds are found to increase monotonically with the size of the world. Second, irrespective of the choice of $R$, we find that $\mathrm{P}_\mathrm{O_2} < \mathrm{P}_c$ is fulfilled only for relatively modest values of $F_\mathrm{XUV}$. More specifically, we require $F_\mathrm{XUV} \lesssim F_\oplus$ to ensure that $\mathrm{P}_\mathrm{O_2}$ stays within the requisite limit(s).

The XUV fluxes of the TRAPPIST-1 planetary system \citep[Table 1]{BEW17}, Proxima b \citep[Table 2]{RBS16}, LHS 1140 b \citep[Section 2]{SBC19}, and the recently discovered TOI-700 d \citep[Section 7.2]{GBS20} are at least an order of magnitude higher than that of Earth. Moreover, their ages are on the order of a few Gyr at the minimum. When both of these results are substituted into (\ref{TotO2gen}), we end up with $\mathrm{P}_\mathrm{O_2}$ that exceeds the desired threshold by a few orders of magnitude. Thus, at least insofar as Earth-like complex life is concerned, these worlds might have a low likelihood of hosting such lifeforms. 

Let us, however, adopt the premise that highly efficient antioxidant mechanisms are functional on worlds with substantial abiotic O$_2$ atmospheres. This brings up an interesting corollary. As the atmospheric O$_2$ inventory increases, so too does the maximal size of organisms that are reliant on ``simple'' mechanisms of O$_2$ capture and transport, with diffusion and blood circulation constituting two common examples. For the latter two cases, the theoretical sizes were estimated in \citet{Alex71} and \citet{CGZM}. Although the scaling laws are independent of the organismal geometry, the prefactors are dependent on it. In view of this fact, by drawing upon \citet[Table 2]{CGZM}, the following heuristic scaling relations are formulated:
\begin{subequations}\label{Rmax}
\begin{eqnarray} \label{Rdiff}
&& \mathcal{L}_\mathrm{diff} \sim  4 \times 10^{-3}\,\mathrm{m}\,\sqrt{\frac{\mathrm{P}_\mathrm{O_2}}{1\,\mathrm{bar}}},\\ \label{Rcirc}
&& \mathcal{L}_\mathrm{circ} \sim 0.1\,\mathrm{m}\,\left(\frac{\mathrm{P}_\mathrm{O_2}}{1\,\mathrm{bar}}\right),
\end{eqnarray}
\end{subequations}
where $\mathcal{L}_\mathrm{diff}$ and $\mathcal{L}_\mathrm{circ}$ denote the maximal sizes of organisms reliant on diffusion and circulation, respectively. We caution that the above scalings represent an oversimplification because physiological parameters such as metabolic rate and permeability of O$_2$ through tissue were held fixed. Based on Fig. \ref{FigPressO}, it is conceivable that atmospheres with $\mathrm{P}_\mathrm{O_2} \sim 100$-$1000$ bar could exist on certain worlds \citep{LB15}. By invoking (\ref{Rmax}) for this range, we obtain $\mathcal{L}_\mathrm{diff} \sim 0.04$-$0.13$ m and $\mathcal{L}_\mathrm{circ} \sim 10$-$100$ m. Thus, even in the absence of sophisticated respiratory systems (e.g., lungs and gills), organisms that are adapted to live in hyperoxic environments might be capable of attaining great sizes. 

Furthermore, in light of the positive scaling between brain mass (and neurons) and body mass \citep{HH16}, worlds with high oxygen inventories may prove to be amenable to the development of large brains and concomitant high intelligence. This notion can be further quantified heuristically in the following fashion. Denoting the brain mass by $\mathcal{M}_b$ and the body mass by $\mathcal{M}$, we use the scaling $\mathcal{M}_b \propto \mathcal{M}^{0.7}$ \citep{HK90,WG10}. Next, we use the fact that metabolic theory and empirical data suggests that neuron mass ($\mathcal{M}_c$) scales as $\mathcal{M}_c \propto \mathcal{M}^{0.25}$ \citep{SAB07}. Hence, the total number of neurons in the brain ($\mathcal{N}_b$) scales as $\mathcal{N}_b \propto \mathcal{M}_b/\mathcal{M}_c \propto \mathcal{M}^{0.45}$. Lastly, we draw upon the geometric scaling $\mathcal{M} \propto \mathcal{L}^3$ (where $\mathcal{L}$ is the body size) to end up with $\mathcal{N}_b \propto \mathcal{L}^{1.35}$. 

Now, let us suppose that the hypothetical organisms under question reach a size that is some fixed fraction of the maximal size. In quantitative terms, this amounts to presuming that $\mathcal{L} \propto \mathcal{L}_\mathrm{diff}$ or $\mathcal{L} \propto \mathcal{L}_\mathrm{circ}$ depending on the context. Of the two categories, organisms with circulatory systems are probably more well-suited for the development of large brains as they enable the efficient delivery of nutrients and oxygen to the brain \citep{SRB04}. Hence, by utilizing (\ref{Rcirc}) in conjunction with the aforementioned scalings, we finally arrive at
\begin{equation}
    \mathcal{N}_b \propto \mathrm{P}_\mathrm{O_2}^{1.35}.
\end{equation}
Therefore, as the scaling with $\mathrm{P}_\mathrm{O_2}$ is quite pronounced, it suggests that the total number of neurons can become quite large in highly oxygenated environments. As the number of neurons is a reasonable proxy for intelligence \citep{HH16}, worlds with high O$_2$ inventories (at least up to a certain limit) might provide conducive environments for the advent of organisms with sophisticated cognitive abilities. 

\section{Discussion and Conclusions}\label{SecConc}
In this work, we sought to investigate the constraints placed by oxygen toxicity on putative Earth-like complex lifeforms on other worlds. We began with a brief summary of how oxygen becomes toxic to both prokaryotes and eukaryotes at high concentrations, and described tentative limits on the maximum partial pressures of O$_2$ ($\mathrm{P}_c$) that are tolerable by Earth's biota. 

Next, we described the various sinks and sources of O$_2$ and employed them to construct a dynamical model for the evolution of atmospheric O$_2$ over time. By imposing the condition $\mathrm{P}_\mathrm{O_2} < \mathrm{P}_c$ must hold true, we studied the prospects for complex life in Figs. \ref{FigOxyTox} and \ref{FigPressO}. One of the major results from the analysis was that worlds that receive temporally averaged XUV fluxes that are an order of magnitude more than Earth are potentially likely to exceed the limits of oxygen toxicity for Earth-based life. Hence, this casts doubt on whether M-dwarf exoplanets such as Proxima b and the TRAPPIST-1 system could host Earth-like complex life. In the event that complex life can survive in hyperoxic environments, we estimated the maximal sizes attainable by organisms and showed that these limits become very large with respect to Earth.

The hypothesis presented in this paper is complementary to that espoused in \citet{SRO19} in two respects. First, \citet{SRO19} showed that CO$_2$ toxicity may move the outer edge of the HZ inward, as higher CO$_2$ concentrations can prevail at the outer edge, whereas this work suggests that O$_2$ toxicity might shift the inner edge of the HZ outward due to the higher XUV fluxes and abiotic buildup of O$_2$. Second, the analysis by \citet{SRO19} indicates that the accumulation of CO (which is toxic to many metazoans on Earth) in atmospheres of M-dwarf exoplanets may render them uninhabitable, while this paper implies that exoplanets around M-dwarfs could become inhospitable to Earth-like biota because of their high XUV fluxes, which aid in abiotic O$_2$ buildup and cause O$_2$ toxicity in the process. Therefore, in a nutshell, CO$_2$, CO and O$_2$ toxicity might jointly conspire to preclude complex Earth-like life on many M-dwarf exoplanets.

Some of the major caveats concerning our analysis are worth pointing out here. First, as noted previously, we made the simplifying assumption that the threshold $\mathrm{P}_c$ as well as the various sources and sinks of oxygen resemble those of modern Earth and are roughly constant over time. Second, our work does not take nonlinear feedbacks for O$_2$ regulation - such as fires and their capacity to destroy organic matter \citep{KK05} - into account. Third, the absorption of O$_2$ by magma oceans especially during the pre-main-sequence phase \citep{WSF18}, and elimination of oxygen by NO$_x$ species produced during lightning \citep{HFH18} are two potentially important effects not incorporated in our simple model, both of which might suppress the buildup of abiotic O$_2$. On the other hand, high fluxes of stellar energetic particles generated by active stars could facilitate the accumulation of O$_2$ \citep{LDF18}, which is also not addressed herein.

On account of the numerous simplifications involved, we must ask ourselves whether the model yields predictions that are readily testable (or falsifiable). The answer is seemingly in the affirmative as described below. For starters, recall that this model suggests that worlds with $\mathrm{P}_\mathrm{O_2} > \mathrm{P}_c$ would not host complex life and vice-versa. The identification of substantial abiotic O$_2$ is feasible through a number of avenues, of which two of the best known metrics are dimer O$_2$-O$_2$ collisional absorption bands at visible wavelengths and CO features in the mid-infrared \citep{MMC14,SMD16}. The presence of complex life may be inferred through a number of avenues ranging from the vegetation red edge \citep{SKP18} and phase-angle-dependent reflectance \citep{DW10} to atmospheric seasonality \citep{OSR18} and transient oceanic phenomena \citep{ML18}.

Hence, in the event that the biosignatures are discovered simultaneously in the \emph{absence} of markers for high abiotic O$_2$, it would lend credence to the notion that relatively modest oxygen levels place a vital constraint on complex life. In contrast, if biosignatures of complex life are discovered concomitantly with massive inventories of abiotic O$_2$, it would imply that oxygen toxicity is not an innate limiting factor for complex life, \emph{contra} the hypothesis espoused in this paper. Therefore, by carefully distinguishing between the various scenarios using appropriate data from future telescopes, it might be possible to gauge whether excessive O$_2$ constitutes a universal bane for complex multicellular life. 

\acknowledgments
It is a pleasure to thank the reviewer, Stephanie Olson, for her perspicacious report that was very helpful in terms of improving the paper. This research was supported by the Department of Aerospace, Physics and Space Sciences at the Florida Institute of Technology.

%\bibliographystyle{aasjournal}
%\bibliography{O2Tox}

\end{document}